# PreMevE-MEO: Predicting Ultra-relativistic Electrons Using Observations from GPS Satellites

**Yinan Feng[1,2], Yue Chen[1], and Youzuo Lin[1,2]**

[1]Los Alamos National Laboratory, Los Alamos, NM, USA.

[2]The University of North Carolina at Chapel Hill, Chapel Hill, NC, USA.

Corresponding author: Y. Chen (cheny@lanl.gov)

**Key Points:**

- Updated PREdictive MEV Electron (PreMevE) model is mainly driven by electron observations from GPS satellites in medium-Earth-orbits (MEOs)

- An innovative machine-learning algorithm combining convolutional neural networks with transformers is developed, optimized, and tested

- New PreMevE-MEO model makes hourly nowcasts of ≥2 MeV electrons inside Earth's outer radiation belt with high fidelity

**Abstract**

Ultra-relativistic electrons with energies greater than or equal to two megaelectron-volt (MeV) pose a major radiation threat to spaceborne electronics, and thus specifying those highly energetic electrons has a significant meaning to space weather communities. Here we report the latest progress in developing our predictive model for MeV electrons in the outer radiation belt. The new version, primarily driven by electron measurements made along medium-Earth-orbits (MEO), is called PREdictive MEV Electron (PreMevE)-MEO model that nowcasts ultra-relativistic electron flux distributions across the whole outer belt. Model inputs include >2 MeV electron fluxes observed in MEOs by a fleet of GPS satellites as well as electrons measured by one Los Alamos satellite in the geosynchronous orbit. We developed an innovative Sparse Multi-Inputs Latent Ensemble NETwork (SmileNet) which combines convolutional neural networks with transformers, and we used long-term in-situ electron data from NASA's Van Allen Probes mission to train, validate, optimize, and test the model. It is shown that PreMevE-MEO can provide hourly nowcasts with high model performance efficiency and high correlation with observations. This prototype PreMevE-MEO model demonstrates the feasibility of making high-fidelity predictions driven by observations from longstanding space infrastructure in MEO, thus has great potential of growing into an invaluable space weather operational warning tool.

**Plain Language Summary**

Electrons traveling nearly light speed inside Earth's outer Van Allen belt have high penetrating ability, and thus pose a major radiation threat to man-made satellites by causing malfunctions of space-borne electronics. Therefore, predicting those ultra-relativistic electrons is significant to all space sectors. Here we update our latest development of a predictive model for megaelectron-volt (MeV) electrons inside the Earth's outer radiation belt, using satellite observations from medium-Earth-orbits (MEO). This new model, called PREdictive MEV Electron (PreMevE)-MEO, focuses on nowcasting ultra-relativistic electron flux distributions across the outer radiation belt, with no need for local measurements of the whole population of trapped MeV electrons except at the geosynchronous orbit (GEO). Model inputs include electrons observed in MEO by up to twelve GPS satellites as well as in GEO by one Los Alamos satellite. We developed an innovative machine-learning algorithm, trained and evaluated a list of models using electron data from NASA's Van Allen Probes mission, and successfully demonstrated the high performance of PreMevE-MEO model. This new model provides hourly nowcasts of incoming nearly light-speed electrons with high statistical fidelity, and thus can make an invaluable tool to space communities.

**1 Introduction**

Solar eruptions often disrupt the space environment near the Earth by sending new charged particles into the region and causing severe local space weather phenomena. For example, southward interplanetary magnetic field often trigger magnetospheric substorms that inject solar wind electrons from the Earth's magnetosphere tail into the near-Earth region inside the geosynchronous (GEO) orbit. When a concert of acceleration and transport processes coherently work out, sometimes those 10s keV electrons can be energized to several megaelectron-volt (MeV) with their intensities reaching sustaining high levels, manifested as MeV electron events. In addition to significantly increasing ionizing dose rates, these ultra-relativistic electrons—with energies greater than or equal to two MeV—are highly penetrating and notorious in internally charging dielectrics on board satellites [Wang et al., 2018]. Repetitive internal chargings over a short time may end up with sudden electrostatic discharging arcs that can lead to subsystem anomalies and even knock out satellites. Therefore, reliably predicting high-intensity events of ultra-relativistic electrons inside the outer radiation belt is of crucial need for space communities.

Routine monitoring of MeV electron events is currently confined to the GEO orbit by several National Oceanic and Atmospheric Administration (NOAA) satellites, whose observations also enable warnings through an operational Relativistic Electron Forecast Model (REFM). For predictions at non-GEO orbits, besides relying upon first-principles models primarily based on the theoretic diffusive framework, a recent new trend is to develop machine-learning models driven by precipitating low-energy electrons observed by Polar Operational Environmental Satellite(s) (POES) in low-Earth-orbits (LEO)—another NOAA satellite missions, taking advantage of the cross-dimensional coherence of electron populations inside the inner magnetosphere as well as the short POES orbital periods. This new idea was first proposed by Chen *et al.* (2016) and Chen *et al.* (2019) successfully constructed the first PREdictive MEV Electron (PreMevE) model based on simple linear predictive filters. Later, Pires de Lima *et al.* (2020) introduced machine-learning (ML) techniques and advanced the model to PreMevE 2.0 making 1- and 2-day forecasts of 1 MeV electrons in the whole outer-belt, and then Sinha *et al.* (2022) further developed PreMevE-2E that expands the forecasts to electrons with ≥ 2 MeV energies, a population with high beta ratio values > ~0.98 and thus also called ultra-relativistic electrons. In a similar vein, the Specifying High-altitude Electrons using Low-altitude LEO Systems (SHELLS) model was first developed by Claudepierre & O'Brien (2020) and recently updated by Boyd *et al.* (2023), which nowcasts 30 keV - 3 MeV electrons with a 1-min time resolution using POES fluxes and Kp index as inputs.

Both PreMevE and SHELLS models characterize MeV electrons inside the outer radiation belt with no requirement of in-situ measurements, however one potential weakness of both lies in the model inputs needed from POES electron measurements in LEO. The current POES satellites, which provide key space weather observations including precipitating electrons, may reach their end-of-life in the next several years and no similar space weather instruments are in plan for the new-generation POES satellites. Therefore, the anticipated discontinuity of POES inputs calls for exploring alternative inputs for PreMevE from other longstanding space infrastructure, and one such candidate is the electron measurements made in medium-Earth-orbits (MEO). Additionally, changing model inputs from LEO to other orbits may necessitate overhauling the model by introducing new ML algorithm(s).

Purpose of this paper is to report how PreMevE has been upgraded to make predictions of ultra-relativistic electron flux distributions across the outer radiation belt using GPS particle data as primary inputs. The updated model, named PreMevE-MEO where the suffix highlights the change of input orbits, ingests sparse input data and provides hourly nowcasts of MeV electron events, incorporating observations from Los Alamos National Laboratory (LANL) space weather detectors aboard GPS satellites in MEO and a geosynchronous satellite in GEO. Still, with no need of in situ measurements of the whole trapped electron population except for at GEO, this new version has demonstrated its capability of meeting predictive requirements for penetrating outer-belt electrons. In the next section, data and model parameters used for this study are briefly described. Section 3 explains in detail our newly designed ML algorithm and how the model is trained, validated, and tested, and Section 4 presents some discussions. This work is concluded by Section 5 with a summary of our findings and possible future directions.

**2 Data and Input Parameters**

Data used in this work include electron observations made by particle instruments aboard one NASA's Van Allen Probes spacecraft (RBSP-a), one LANL GEO satellite, and twelve GPS satellites over a period ranging from February 2013 to June 2018, as shown in Figure 1 left panels. Ultra-relativistic electron spin-averaged flux distributions are in-situ measured near the geomagnetic equatorial plane by Relativistic Electron-proton Telescope (REPT, Baker *et al.* [2012]) experiment aboard RBSP-a spacecraft at L ≤ 6, and by Energy Spectrometer for Particles (ESP, Meier *et al.* [1996]) instrument carried by one LANL GEO satellite at L ~ 6.6. As the overview presented in Figure 1A1, hourly-binned integral fluxes of ≥2 MeV electrons are the target dataset that is a function of L-shell over the 1957-day interval. Here we use McIlwain's L values (McIlwain, 1966) calculated from the quiet Olson and Pfitzer magnetic field model (Olson & Pfitzer, 1977) together with the International Geomagnetic Reference Field model. These ≥2 MeV electron target data are used for model training, validation, and test, and are not needed as model input (except for at GEO) for making predictions. Target data over a selected seven days are replotted in Figure 1A2 to show details, from which L-shell traces of both RBSP-a satellite (with a ~9 hr orbital period) and the LANL GEO satellite (24 hr period with diurnal variations of L-shell locations) can be clearly seen.

Electron fluxes measured by GPS satellites are the primary model input in this study. The GPS orbit is a typical circular MEO with an altitude ~20,200 km and nominal inclination of 55º, crossing the heart of electron outer radiation belt (L~ 4.2) four times daily. GPS constellation has a long history of carrying space particle instruments developed by LANL since the 1980s [Cayton *et al.*, 1992]. Recently, LANL has released decades of GPS particle data to the public [Morley *et al.*, 2017] as part of national efforts of enhancing space-weather preparedness, and those data (including version v1.08 used in this work) can be retrieved from https://www.ngdc.noaa.gov/stp/space-weather/satellite-data/satellite-systems/gps/. GPS electron fluxes used here are from the latest generation of LANL charged particle instruments called Combined X-ray Dosimeter (CXD, Tuszewski *et al.* [2004]), which measure electrons with energies from 120 keV to more than 5 MeV. This study uses ≥2 MeV data from twelve GPS satellites as model inputs, and example data from four GPS satellites ns53, ns57, ns59 and ns66 are presented in Figure 1 B-E (other GPS satellite data are similar and not shown here). Due to the GPS orbit inclination, measured flux values depend on both L-shell—the whole trapped MeV population are only measurable at equator crossings—and geomagnetic activity levels. Intensity increments in MeV electron events can be seen from GPS data in Panels B1-E1, coincidental to the increments observed by RBSP-a in Panel A1 particularly in the high intensity regions (with red color) at L-shells between ~ 4 - 6. Because of the longer periods of MEO than LEO, data from one single GPS satellite is both temporally and spatial sparse for an hourly resolution, as shown in Figure 1 right panels, and thus we need multiple GPS satellites (up to twelve) for better input coverage. Indeed, developing new ML algorithms to appropriately treat the sparseness in both target (Panel A2) and input data is one major challenge of this work. Another limitation of GPS data is the lack of coverage at L< ~4.

All RBSP-a, LANL GEO and GPS electron fluxes in Figure 1 are binned for 1 hour time resolution, and the L-shell bin size is 0.2. Throughout this work, logarithmic values of ≥2 MeV electron fluxes from GPS and GEO satellites form the input datasets, or the predictors, being used to predict the logarithm of ≥2 MeV trapped electron flux distributions, also referred to as "target". For model development, 70% of all data (over 1369 days starting from February 20$^{th}$, 2013) are used as the training set, 20% of data (392 days from November 20$^{th}$, 2016) are used for validation and hyperparameter optimization, and the remaining 10% (196 days starting from December 17$^{th}$, 2017) are kept for final test. For each individual satellite, electron fluxes across all L-shells are standardized via subtracting the mean and dividing by the standard deviation after logarithm (both the mean and standard variation are computed with the training set). L-shell coverage of this model is confined to 3.0–7.2 for this study. Additional input includes the low-L boundary locations, which can be identified from POES NOAA-15 electron measurements to inform model the inner boundary locations of ultra-relativistic electrons. More details are described in the Appendix.

## 3 PreMevE-MEO: Model Construction and Performance

### 3.1 SmileNet and Model Specifications

For PreMevE-MEO, we designed a new algorithm called Sparse Multi-Inputs Latent Ensemble Network (SmileNet) that excels in processing sparse and multi-source input datasets and handles long-range correspondences in data with mask-aware partial convolution and Transformer techniques. As outlined in Figure 2, SmileNet includes two sequential principal components, the Embedding Network and the Processing Network, and each serves distinct functions in data handling and analysis: the Embedding Network embeds sparse data from each individual satellite into a list of unified latent representations which are then ensembled and averaged, and the Processing Network aims to capture the long-range correspondences and complex patterns in the embedding latent representation and generate the dense output.

Inside SmileNet, the Embedding Network is constructed with a series of subnetworks, each receiving electron data from a single satellite with the consideration of not perfect inter-calibrations between individual instruments. Each subnetwork is a convolutional neural network (CNN) with residual connections and a partial convolution layer. Here the convolutional layer embeds input data into a set of more abstract feature maps, called latent representations. Each feature map is generated by sliding a learnable window filter across the input data to capture unique features and patterns. The partial convolutional layer comes from advanced techniques in image inpainting [Liu *et al.*, 2018; Li *et al.*, 2022]. This layer is adept at processing incomplete input data alongside corresponding masks, effectively managing the intricacies of satellite data with varying completeness. Mathematically, the partial convolution at every location is defined as $x' = \begin{cases} W^T(X \odot M)\frac{n}{sum(M)} + b, & if\ sum(M) > 0, \\ 0, & else, \end{cases}$ where $W$ and $b$ be the convolution filter's weights and bias, $X$ are the input values for the current convolution layer, $M$ is the corresponding binary mask to indicate whether a pixel (or the input at a L-Time bin) is valid or not, $\odot$ is element-wise multiplication, and $n$ is the number of pixels in $X$. After each partial convolution, the masks are checked and updated: if at least one valid input value in a window, the output position is updated to be valid; otherwise, they remain invalid. The residual connections in neural networks are shortcuts that skip one or more layers, directly passing the input to deeper layers and helping prevent the vanishing gradient problem during training.

In comparison, the Processing Network is constructed as a Vision Transformer network (ViT) [Dosovitskiy *et al.*, 2020] with positional embeddings. Based on experience, we chose a two-layer network architecture to have a model with a reasonable size to effectively tackle the problem's complexity. ViT is a pioneering model that adapts the transformer architecture, originally designed for natural language processing, to image processing tasks. Instead of utilizing convolutional operations, ViT segments an input image (or equivalently input electron flux distributions in the L-Time 2D space) into a grid of smaller, fixed-size pieces known as patches. Each patch serves as a discrete input unit for the model, which is transformed into a vector via a linear projection before being fed to transformer layers. This approach enables capturing both local and global information from the image in a structured manner. Position embeddings [Vaswani *et al.*, 2017] are added to the patches before processing to ensure the model maintains an awareness of the original positioning of each patch in the image. Sinusoidal positional embeddings are used in this study, which are pre-defined sine/cosine functions of the token position and the number of dimensions of feature maps.

Figure 2 illustrates how the transformer layer functions. The architecture of a transformer layer starts with a normalization process to stabilize data scale and enhance processing efficiency. Following normalization is a multi-head attention layer that is the key component. It allows the model to weigh the importance of all other patches related to a specific patch and construct a global understanding of the visual input, which is especially effective for identifying patterns and relationships in data. Post multi-head attention, the processed information is once again normalized, ensuring that the output maintains a consistent scale for further processing. The final step in the transformer layer is a multi-layer perceptron (MLP) block. This MLP block refines the data, extracting more complex patterns and relationships, thus deepening the model's understanding of the input.

In this way, Processing Network digests the latent representations derived by Embedding Network and employs Transformers to effectively interpret spatial patterns by establishing long-range correspondences. Afterwards, each data patch is unpatched to its original dimension, and there is a 1×1 convolution layer that maps the number of feature map to 1, i.e., mapping the representation of data into the target output shape).

PreMevE-MEO covers L-shells from 3 to 7.2 with a grid size of 0.2 and has a time step length of one hour. When preparing model input, GEO electron fluxes are directly combined with GPS fluxes and both are binned accordingly, while the POES low-L boundary locations are transformed into a format akin to satellite data. Specifically, we assigned a flux value of $3x10^3/cm^2/s/sr$ to the low-L boundary L-shell locations, and thus indeed treating the transformed boundary data as an additional pseudo satellite fluxes measured at L-shells much lower than GPS coverage. This simplified treatment is justified by observations as explained in Appendix as well as by performance of the model. At each 1-hr time step, our model utilizes L-sorted electron fluxes from preceding 72 hours as input, with the objective of generating predictions that fully cover the outer belt for corresponding time bins.

Given significant sparsity in both input and target data, at each time step we let model output data corresponding to the same 72-hr bins to ensure sufficient signal for model training, and predictions in the latest hour bin are the nowcast results from the model. Consequently, the training loss is calculated over these 72 hours to model optimization. In this study, model performance is primarily measured by performance efficiency (PE) at each individual L-shell at datapoints with target observations available. PE quantifies the accuracy of predictions by comparing to the variance of the target, defined as $PE = 1 - \sum_{j=1}^{M}(y_j - f_j)^2 / \sum_{j=1}^{M}(y_j - \bar{y})^2$, where the observation *y* and model prediction *f* are both with size *M*, and $\bar{y}$ is the mean of *y*. PE's highest score 1.0 indicates a perfect model with all predicted values exactly matching observed data, a zero PE suggests a performance equivalent to a climatology model, and a negative PE means poor model performance.

**3.2 PreMevE-MEO Performance in Nowcasting ≥2 MeV Electron Distributions**

Based upon SmileNet, a list of models has been trained and tested with varying hyper-parameters and input combinations. The top performer among them is named PreMevE-MEO, which takes the data from all twelve GPS satellites, the LANL GEO, and the low-L boundary as inputs to generate nowcasts at L-shells between 3 and 7.2. To improve performance at low L-shells, a post-processing is introduced at each time step where the outputs at L grids within [3.0, 4.0) from the ML model are replaced by linear interpolation results between ML model predictions at higher L-shells and the threshold flux value at the low-L boundary location.

Figure 3 presents an overview of PreMevE-MEO results and performance. Model nowcasts over the whole 1957-day interval are plotted in Panel B, showing highly similar dynamics compared to the target in Panel A. Indeed, most MeV electron events are well captured in terms of both intensities and L-Time shapes of the areas filled by red and yellow colors; exceptions exist mainly for significant electron dropouts, e.g., the one in the test interval on days ~1800. Panel C presents model nowcasts with no post-processing at low L-shells, and electron dynamics at L below 4—e.g., the borderline of green color at L ~3—are not as well predicted during the out-of-sample interval (including both validation and test) as in Panel B, particularly missing the major deep electron injection event starting at day 1661. Panel D further plots PreMevE-MEO PE values in orange color for out-of-sample days as a function of L-shell. Comparing to the PE curve in blue with no post-processing, the significant improvement at L-shells below 3.5 highlights the necessity of using low-L boundaries for post-processing. The orange PE curve has its PE value of 0.92 at $L$ = 4.2 and decreasing in both directions with a minimum value of 0.50 at L = 3.0, which is reasonable because GPS satellites observe the whole trapped electron populations only at equatorial crossings (L ~ 4.2). PE values at GEO (L > 6.2) are also higher than ~0.85 because of the dual roles played by LANL GEO data, and the gap in PE curve at L = 6.2 is due to the removal of RBSP-a data at L-shells above 6. Here PreMevE-MEO has a mean PE value of 0.83, which is higher than our previous nowcasting model using POES data [Chen et al., 2019], and most importantly, the PE values are all higher than 0.82 at L≥4 where majority of ultra-relativistic electrons reside.

Focusing on the out-of-sample interval, Figure 4 replot the target flux distributions in Panel A1, PreMevE-MEO predictions in Panel B1, and their ratios in $\log_{10}$ scale in Panel C1 for the 588 days of validation and test periods, as well as for a selected seven days in the right panels. In Panels C1 and C2, error ratios are dominated by the green color, indicating ratios value close to one and thus great agreement between target and prediction, while red (purple) means over- (under-) predictions. In Panel C1, it is visible that over-predictions (in red) are mostly at L-shells below 3.5, except for the one major loss events starting on ~ day 1774 in which over-predictions extend to higher L-shells, while under-predictions (purple) are mostly confined to Lshells below 4. This is not surprising because the current PreMevE-MEO has no inputs at low L-shells except low-L boundaries, and it is possible that adding more restraining inputs from LEO may improve model performance at low L-shells.

Even more details of PreMevE-MEO predictions are presented in Figures 5 and 6. Figure 5 shows that, at five individual L-shells over 180 days in the test period, PreMevE-MEO predictions (in red) closely track most of the ups and downs of ≥2 MeV electron fluxes as observed (black), except for the enduring major electron dropout observed at L=3.4. As one example, Figure 6 plots at L=5.6 how PreMevE predictions closely track observed increments and decays in MeV electron events across the whole validation and test periods with a PE value of 0.88. All these demonstrate the reliability of PreMevE-MEO model.

**4 Discussions**

Besides the PreMevE-MEO model as discussed above, we have done a long list of experiments, ranging from model ablation tests to hyper-parameter adjustments. Here we present a brief discussion to highlight some results to help understand our model's design choice.

First, we experimented to compare SmileNet with two other basic network architectures as baselines: One only uses the CNN embedding network (without the Transformer layer), and the other is a simple 10-layer MLP network with Gaussian Error Linear Unit (GELU, Hendrycks *et al.* [2016]) as the activation function. As in Figure 7, the PE curve for CNN only model (the yellow-green curve) has much lower values than PreMevE-MEO (thick orange) except for at L ~ 4.4, which underscore the necessity of our two-stage network architecture. A CNN alone network is inadequate for capturing the complex patterns and long-range correspondences present in this study. Similarly, the curve for MLP model (the cyan curve) have even lower PE values, primarily due to its inability to handle sparse data effectively.

Next, we evaluated the input influence of using a different number of GPS satellites. Instead of utilizing input data from all satellites, we retrained models with subsets of four and eight GPS satellites. For example, here we grouped input data into three sets of four satellites, incorporating scenarios where the four GPS satellites either shared the same orbital plane or did not; and one set of eight satellites. As shown in Figure 7, after all models being retrained, the PE curves for three four-satellite groups and one eight-satellite group all cluster together with their values slightly lower than those of PreMevE-MEO (except for the GEO region). It is interesting to see the brown curve for the group of four GPS satellites in three orbital plans have slightly higher PE values than other subsets at L > 4.2, while the purple curve for the eight-satellite group have lower PE values than others in the same region. One possible explanation is the varying data quality for different satellites, although further investigation is desired. Even with noticeable changes, those clustered PE curves suggests the robustness of our model: reducing the number of satellites for inputs only slightly impacts model performance.

We further evaluated and demonstrated our model's robustness by varying the number of available GPS satellites for inputs without retraining the model. Specifically, after PreMevE-MEO being successfully trained with data from all 12 satellites, the same model was tested using inputs from only eight satellites (the same set as in previous experiments). Its PE curve, the dashed orange curve named PreMevE-MEO (8 sat) in Figure 7, demonstrates the model's commendable performance with only a slight degrade compared to the original model. These results have practical significance because it means the current PreMevE-MEO model nowcasts will continue to be reliable even with the inputs from one or several GPS satellites being discontinued in the future.

Besides PE, the model's performance can also be measured from scatter plots comparing predictions to target over the entire out-of-sample interval, as shown in Figure 8 for three models. In each 2D histogram, the position of each pixel is determined by predicted and target flux values, and the pixel color counts the occurrences over the pixel bin. In each panel, the diagonal indicates a perfect match, and the dark gray (light gray) dashed lines on both sides mark error ratios of 3 and 1/3 (5 and 1/5) between predicted and observed (original flux values not in logarithm). For all three models, majority of the points (in red) fall close to the diagonal and are well contained between the two factor-3 lines, particularly the points in the upper right quarter corresponding to high intensities during MeV electron events. The two percentages in the lower right show the portions of data points that fall with the two pairs of factor lines, and the red number in the second and third row is the correlation coefficient (CC) value. For example, PreMevE-MEO in Panel A has a high CC value of 0.95, and about 87% (94%) output have error ratios within the factors of 3 (5). Other models have similar numbers. In addition to PE, all these CC values and percentages help to further quantify the confidence of model outputs.

This feasibility study has demonstrated that inputs from GPS satellites in MEO can be used as primary inputs to predict outer-belt electron dynamics, and there are future directions planned for the next. First, in this study POES data are purposely filtered to only provide low level dynamic information, that is the low-L boundary locations, for two reasons: One is to highlight the role of GPS data in MEO as inputs, and two is to test whether alternative LEO inputs can be used because low-L boundary locations could be identified from similar data measured by other LEO satellites. Since current results suggest that the low-L boundary is helpful but not ideal, one reasonable next step is to combine POES data in LEO, GPS data in MEO, and LANL data in GEO as inputs to improve model performance over a wider L range than this prototype PreMevE-MEO model. Second, we also plan to apply the same model framework to make forecasts of electron distributions as our previous PreMevE models, by including electrons at other energies and/or upstream solar wind conditions. Last, model performance eventually depends on the quality of the training data, and therefore improving the agreement between RBSP-a and LANL GEO data is also one future direction.

Finally, the newly designed SmileNet excels in its flexibility and strength in handling sparse inputs from various data sources. It's proven to be highly adaptable to manage time-series data with different number of satellites with multiple energy channels, which makes it useful for various space weather applications. The natural input and output of SmileNet are both in a 2D format, which intrinsically suits it well for tasks requiring two-dimensional data, for example, a predictive model for global distributions of substorm injected particles as a function of L-shell and local time. In this work, we demonstrated that SmileNet can effectively learn spatiotemporal dynamics for prediction using sparse data. The challenges and characteristics observed in this study are not unique to our problem and are common across various scientific disciplines. For example, in geophysical applications, the dynamics inherently involve both spatial and temporal components, and data sparsity is a frequent issue. We anticipate that, with some adaptation, SmileNet could address similar challenges in other domains, such as sea temperature modeling, channel turbulence, and multiphase flow.

## 5 Summary and Conclusions

In this work, we have tested (1) different neural network algorithms and (2) different model inputs, with the goal of upgrading our predictive model for ultra-relativistic electrons inside the Earth's outer radiation belt. The updated PreMevE-MEO model is primarily driven by electron data in MEO from up to twelve GPS satellites, as well as data from one Los Alamos satellite in the geosynchronous orbit. An innovative SmileNet algorithm, combining CNN with Transformers to ingest and learn from sparse data with multiple sources, has been designed, trained, optimized, and fully tested. It is shown that PreMevE-MEO can provide hourly nowcasts of ≥2 MeV electron distributions with high fidelity: model PE value of 0.83 averaged for all L-shells with a maximum (minimum) of 0.92 (0.49) at L = 4.2 (3.0) and a high correlation coefficient value of 0.95 to observations. This prototype PreMevE-MEO model demonstrates the feasibility of making high-fidelity predictions driven by observations from longstanding space infrastructure in MEO, thus has great potential of growing into an invaluable space weather operational tool.

**Acknowledgments**

This work was performed under the auspices of the U.S. Department of Energy and supported by the Laboratory Directed Research and Development (LDRD) program (award 20230786ER). We gratefully acknowledge the CXD instrument team at Los Alamos National Laboratory. We also want to acknowledge the PIs and instrument teams of LANL GEO ESP, RBSP REPT, and NOAA POES SEM2 for providing measurements and allowing us to use their data. Thanks to CDAWeb for providing OMNI data.

**Open Research**

RBSP REPT, GPS CXD and NOAA-15 SEM2 data used in this work were downloadable from the missions' public data websites (https://www.rbsp-ect.lanl.gov, https://www.ngdc.noaa.gov/stp/space-weather/satellite-data/satellite-systems/gps/, and http://www.ngdc.noaa.gov), while LANL GEO data are downloadable from OSF repository website at https://doi.org/10.17605/OSF.IO/RCGPW.

A code release request has been filed internally at the corresponding author's institution due to requirements for copyright and export control. This PreMevE-MEO research code will be made publicly available when this study being published.

**Appendix: Identify low-L boundary locations from POES NOAA-15 P6 data**

Due to the limited coverage of GPS data on L-shells, we introduced the low-L boundary locations to help inform the dynamics of ultra-relativistic electrons at low L-shells < ~4. Here the low-L boundary is loosely defined as the inner boundary of the electron outer radiation belt, where ≥2 MeV electrons have a low threshold flux value of ~$3 \times 10^3$ /cm$^{-2}$/s/sr. These low-L boundary values can come from different observational sources but here for simplicity we use the values identified from NOAA-15 observations as shown in Figure A1.

The P6 channel of Space Environment Monitor 2 (SEM2) [*Evans and Greer*, 2000] proton detectors on board POES satellites was original designed to detect multiple-MeV protons, but has been reported to essentially measure ~1 MeV electrons in the outer belt region [*e.g.,* Chen et al., 2019]. In Figure A1, Panel A presents the count rates from the 90° P6 telescope on NOAA-15, binned and plotted as a function of L-shell and time. In each three-hour time bin, the low-L boundary location is identified from the low threshold value, which was selected to be 3 #/s in this study. These identified low-L boundary locations closely track the L-shell locations of ~$3 \times 10^3$ /cm$^{-2}$/s/sr in RBSP REPT ≥2 MeV flux distributions, which can be eyeballed from Panel B. (The same identified locations are overplotted in white in both Panels A and B.) Statistical study shows that the averaged RBSP ≥2 MeV electrons at those POES-derived low-L boundary locations have a mean flux value of $2.8 \times 10^3$ /cm$^{-2}$/s/sr with a standard deviation of $3.1 \times 10^3$ /cm$^{-2}$/s/sr, and the locations identified from POES and the locations identified from RBSP—using the threshold value of $3.0 \times 10^3$ /cm$^{-2}$/s/sr on RBSP ≥2 MeV flux profile in each time bin—have a correlation coefficient of 0.85, and the two identified L locations have a mean absolute difference value of 0.13+/-0.12, which is deemed acceptable for this prototype study considering the model's L-shell grid size of 0.2.

# References

Baker, D. N. et al. (2012). The Relativistic Electron-Proton Telescope (REPT) instrument on board the Radiation Belt Storm Probes (RBSP) spacecraft: Characterization of Earth's radiation belt high-energy particle populations. *Space Sci. Rev.*, 10.1007/s11214-012-9950-9.

Boyd, A.J., Green, J.C., O'Brien, T. P., and Claudepierre, S. G. (2023). Specifying High Altitude Electrons Using Low-Altitude LEO systems: Updates to the SHELLS Model. *Space Weather*, **21**(3), e2022SW003338. https://doi.org/10.1029/2022SW003338

Cayton, T. E., P. R. Higbie, D. M. Drake, R. C. Reedy, D. K. McDaniels, R. D. Belian, S. A. Walker, L. K. Cope, E. Noveroske, and C. L. Baca (1992). BDD-I: An Electron and Proton Dosimeter on the Global Positioning System, Final Report, Tech. Rep., Los Alamos Natl. Lab., LA-12275, https://inis.iaea.org/collection/NCLCollectionStore/_Public/23/071/23071093.pdf

Chen, Y., G. D. Reeves, G. S. Cunningham, R. J. Redmon, and M. G. Henderson (2016), Forecasting and remote sensing outer belt relativistic electrons from low earth orbit: *Geophysical Research Letters*, **43**, 1031–1038

Chen, Y., Reeves, G. D., Fu, X., & Henderson, M. (2019). PreMevE: New predictive model for megaelectron-volt electrons inside Earth's outer radiation belt. *Space Weather*, **17**(3), 438–454.https://doi.org/10.1029/2018SW002095

Claudepierre, S. G., & O'Brien, T. P. (2020). Specifying High-Altitude Electrons Using Low-Altitude LEO Systems: The SHELLS Model. *Space Weather*, **18**(3), e2019SW002402. https://doi.org/10.1029/2019SW002402

Dosovitskiy, A., Beyer, L., Kolesnikov, A., Weissenborn, D., Zhai, X., Unterthiner, T., & et al. (2020). An image is worth 16x16 words: Transformers for image recognition at scale. In Ninth International Conference on Learning Representations.

Evans, D. S., Greer, M. S., & (U.S.), S. E. C (2000). Polar orbiting environmental satellite space environment monitor-2: instrument description and archive data documentation. Boulder, CO: U.S. Dept. of Commerce, National Oceanic and Atmospheric Administration, Oceanic and Atmospheric Research Laboratories, Space Environment Center.

Hendrycks, D., & Gimpel, K. (2016). Gaussian error linear units (gelus). arXiv preprint arXiv:1606.08415.

Liu, G., Reda, F. A., Shih, K. J., Wang, T. C., Tao, A., & Catanzaro, B. (2018). Image inpainting for irregular holes using partial convolutions. In Proceedings of the European conference on computer vision (ECCV) (pp. 85-100).

Li, W., Lin, Z., Zhou, K., Qi, L., Wang, Y., & Jia, J. (2022). Mat: Mask-aware transformer for large hole image inpainting. In Proceedings of the IEEE/CVF conference on computer vision and pattern recognition (pp. 10758-10768).

McIlwain, C. E. (1966). Magnetic coordinates. *Space Science Reviews*, **5**(5), 585–598. https://doi.org/10.1007/BF00167327

Meier, M. M., Belian, R. D., Cayton, T. E., Christensen, R. A., Garcia, B., Grace, K. M., et al. (1996). The energy spectrometer for particles (ESP): Instrument description and orbital performance. In Workshop on the Earth's trapped particle environment (Vol. 383, pp. 203–210). New York: Am. Inst. Phys. Conf. ROC.

Morley, S. K., J. P. Sullivan, M. R. Carver, R. M. Kippen, R. H. W. Friedel, G. D. Reeves, and M. G. Henderson (2017). Energetic Particle Data From the Global Positioning System Constellation, *Space Weather*, 15, 283–289, doi:10.1002/2017SW001604

Olson, W. P., & Pfitzer, K. A. (1977). Magnetospheric magnetic field modeling. Annual scientific report. Huntington Beach.

Pires de Lima, R., Chen, Y., & Lin, Y. (2020). Forecasting megaelectron-volt electrons inside Earth's outer radiation belt: PreMevE 2.0 based on supervised machine learning algorithms. *Space Weather*, **18**, e2019SW002399. https://doi.org/10.1029/2019sw002399

Sinha, S., Chen, Y., Lin, Y., & Pires de Lima, R. (2021). PreMevE update: Forecasting ultra-relativistic electrons inside Earth's outer radiation belt. *Space Weather*, **19**, e2021SW002773. https://doi.org/10.1029/2021SW002773

Tuszewski, M., T. E. Cayton, J. C. Ingraham, and R. M. Kippen (2004). Bremsstrahlung effects in energetic particle detectors, *Space Weather*, **2**, S10S01, doi:10.1029/2003SW000057

Vaswani, A., Shazeer, N., Parmar, N., Uszkoreit, J., Jones, L., Gomez, A. N., ... & Polosukhin, I. (2017). Attention is all you need. *Advances in neural information processing systems*, **30**.

Wang, J.Z. et al. (2018), Electron Environment Characteristics and Internal Charging Evaluation for MEO Satellite, *IEEE Tran. Nucl. Sci.*, Vol **65**, No. 8

# Figures

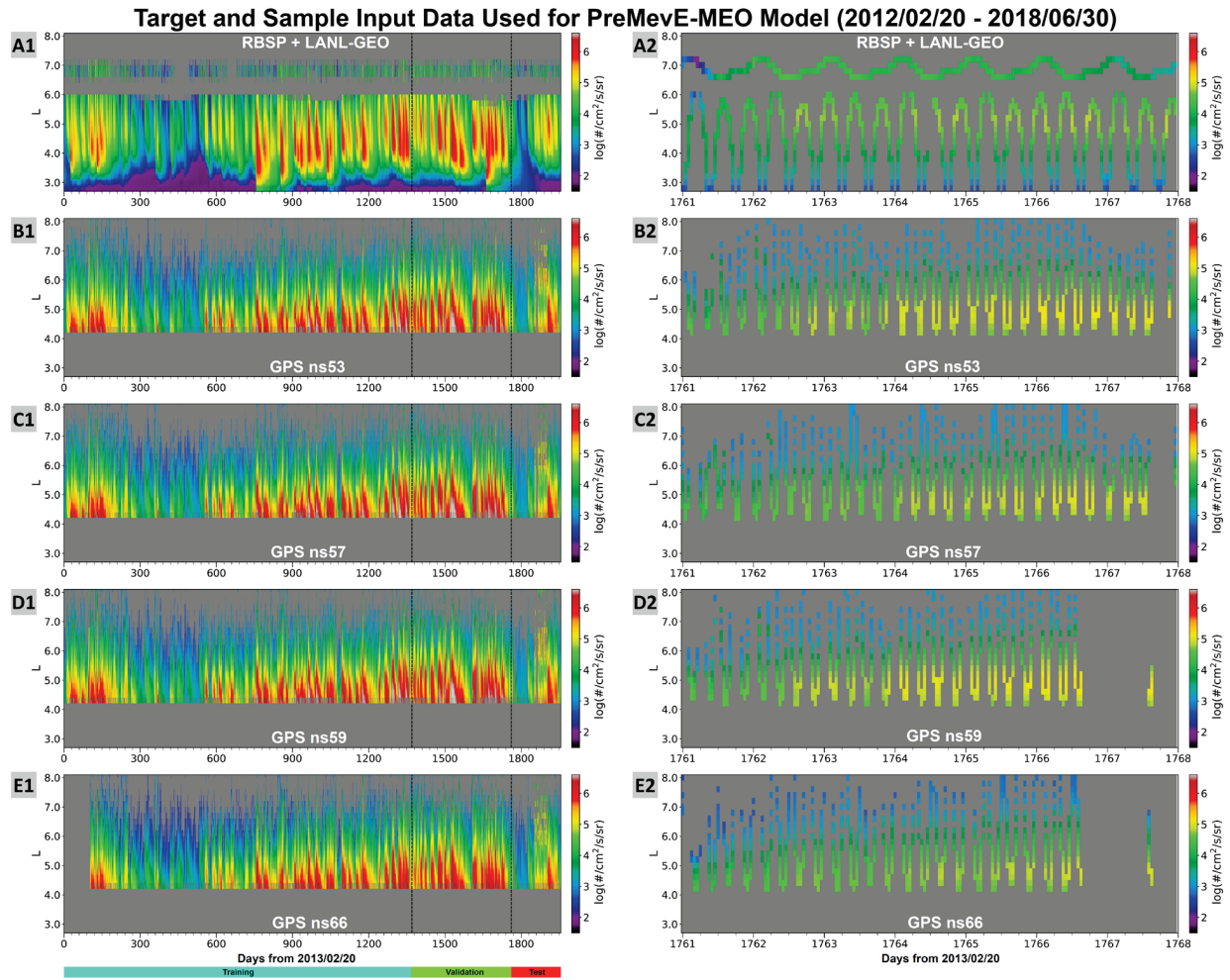

**Figure 1. Overview of electron data used for this study.** Panels in left column present electron fluxes sorted by L-shells and time over the whole 1,957-day interval starting from February 20th, 2013 to June 30th, 2018, and panels in right show hourly-binned fluxes in the same format over seven selected days for details. **A1-A2)** Target flux distributions of ≥2 MeV electrons by RBSP-a at L ≤ 6 and one LANL GEO satellite at L~6.6. **B1-B2)** Input of ≥2 MeV electron fluxes from GPS ns53 satellite. **C1-C2, D1-D2, and E1-E2)** Input electron fluxes from GPS ns57, ns59, and ns66 satellites, respectively. Data in left panels are separated for training, validation, and test, as indicated by the two vertical dashed lines as well as the bottom color bars. Days plotted in right panels are in the beginning of the test interval.

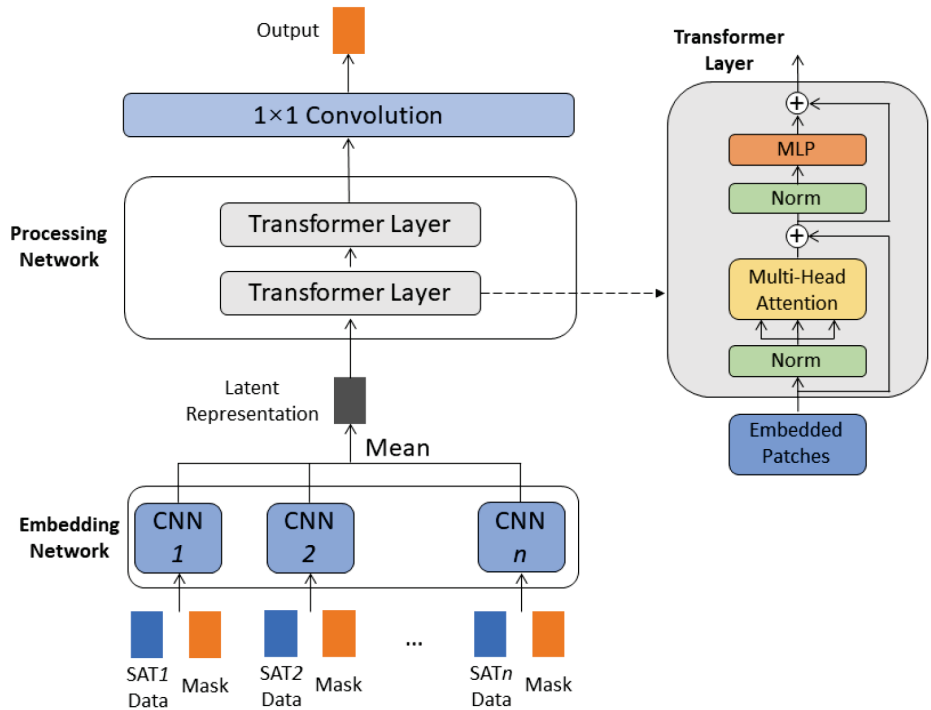

**Figure 2**. **Flowchart of the SmileNet algorithm**. It contains two sequential major components: 1) Embedding Network as the input layer, includes a list of CNN sub-networks used to embed the sparse individual GPS data, and 2) Processing Network is a two-layer transformer to identify long-range correspondences for final output. The final output consists of 1×1 convolution layer.

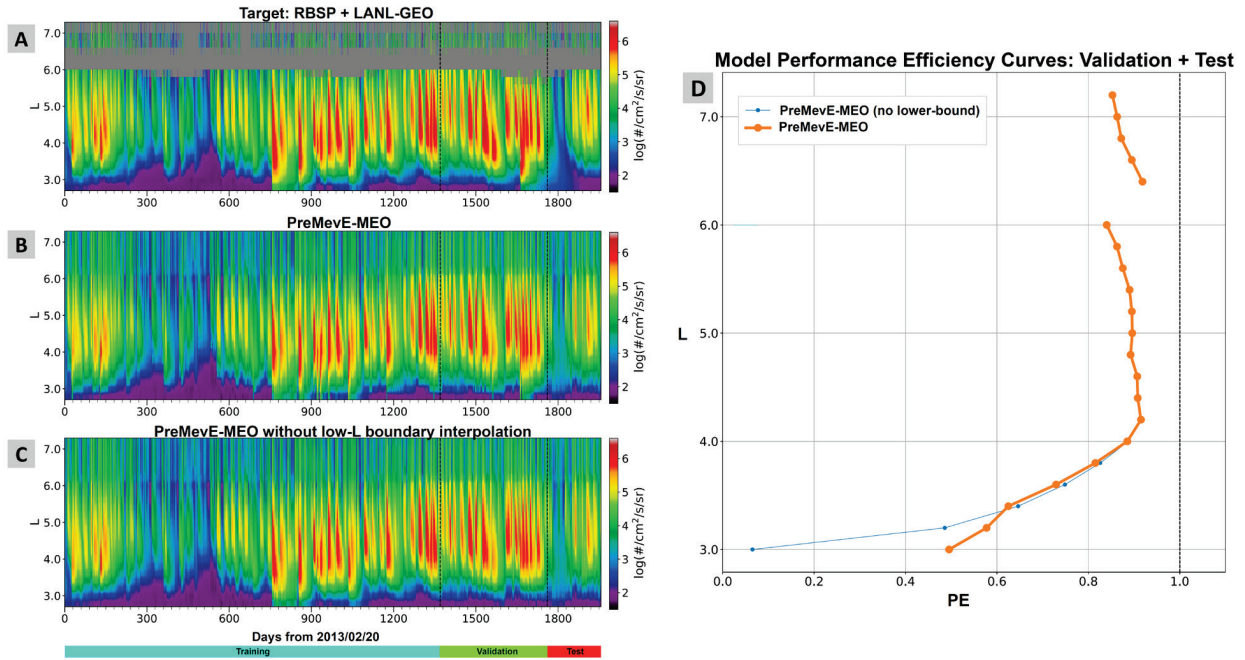

**Figure 3. Summary of PreMevE-MEO model results.** Panel (A) shows observed target electron flux distributions over the whole 1957-day interval, compared to Panel B showing nowcasting predictions from PreMevE-MEO as well as to Panel C model predictions with no low-L boundary post-processing. In Panel D, model performance efficiency (PE) values for the out-of-sample interval (validation and test subsets) are plotted as a function of L: orange (blue) curve for the model with (without) post-processing at low L-shells.

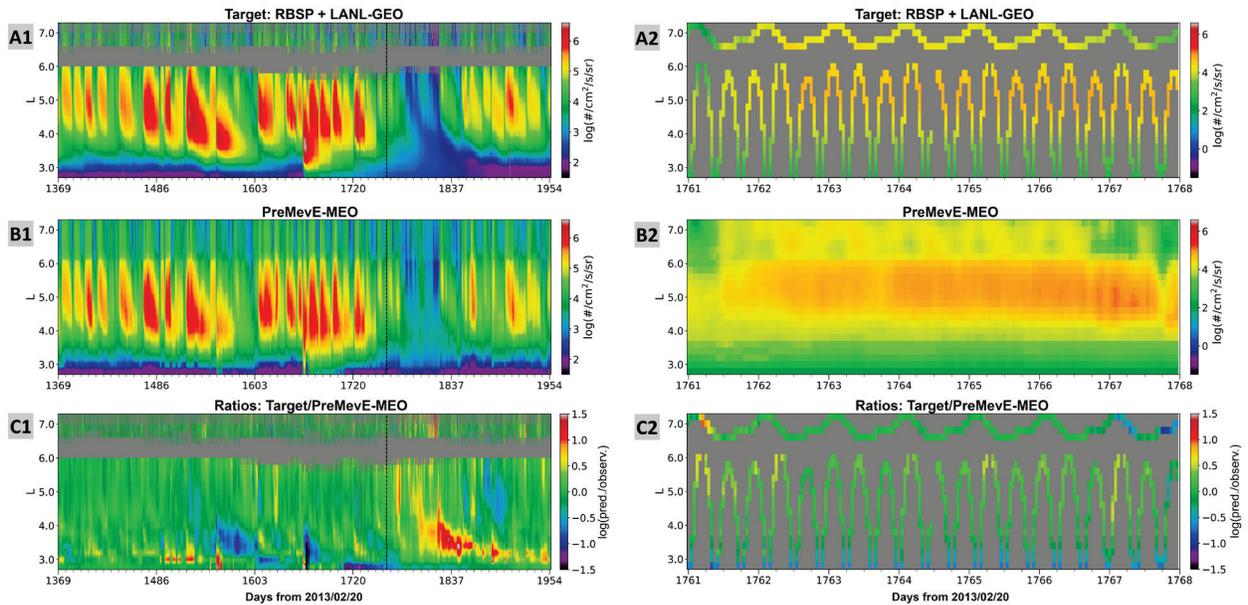

**Figure 4. PreMevE-MEO model predictions over the 588-day out-of-sample interval across all L-shells.** Panel (A1) shows the observed target electron flux distributions over the validation and test intervals, Panel (B1) shows nowcasting predictions from PreMevE-MEO, and Panel (C1) presents error ratio (in $\log_{10}$ scale) distributions between predictions and target data. Panels A2, B2, and C2 plot distributions in the same format over seven selected days showing more details.

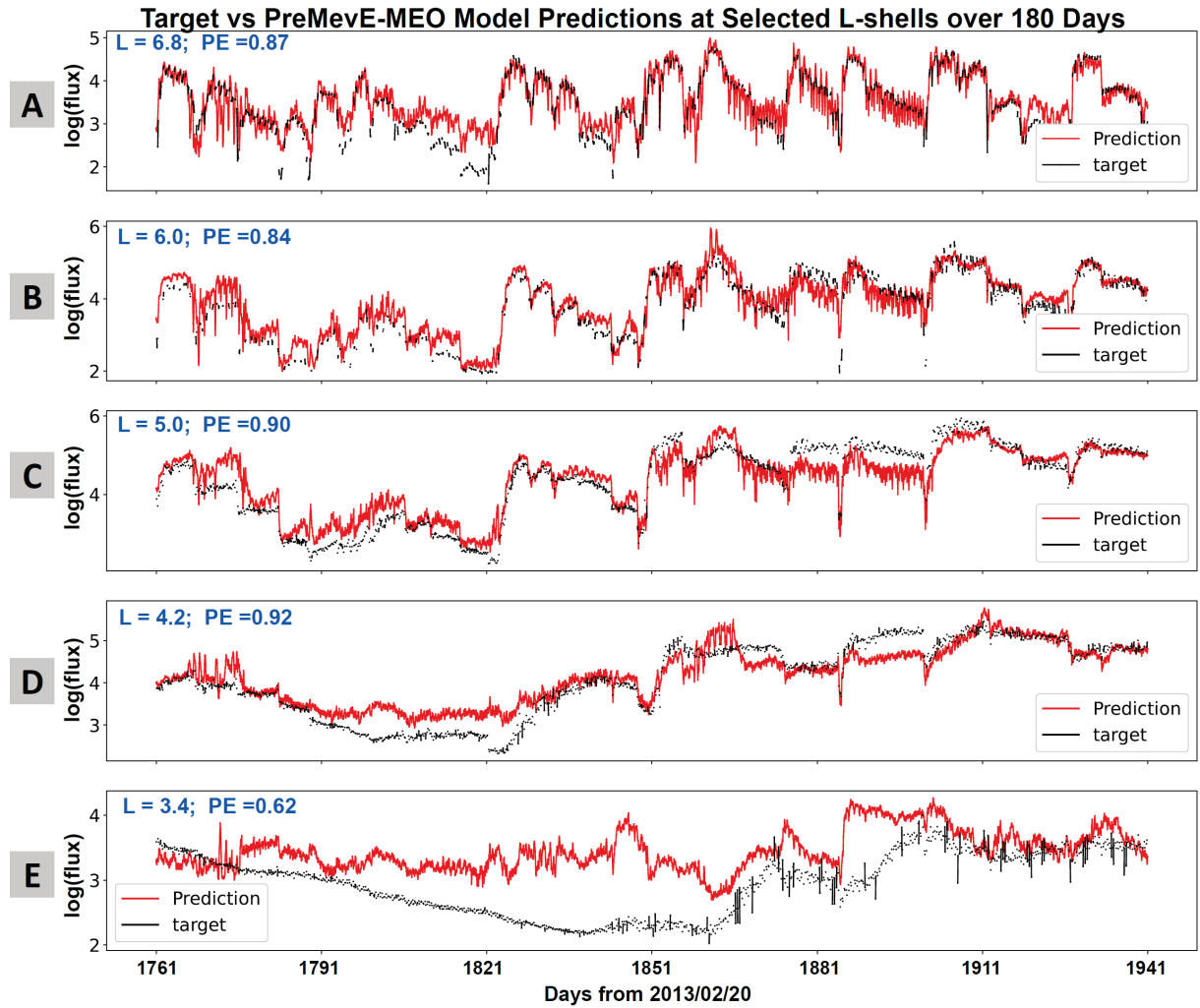

**Figure 5. PreMevE-MEO nowcasting results for ≥ 2 MeV electron fluxes for individual L-shells over 180 days in the test period.** Panels from the top to bottom are for five L-shells at 6.8 (GEO), 6.0, 5.0, 4.2, and 3.4, respectively. In each panel, the target is shown in black, and the predictions are shown in bright red color. The PE value shown for each L-shell is calculated for all validation and test days.

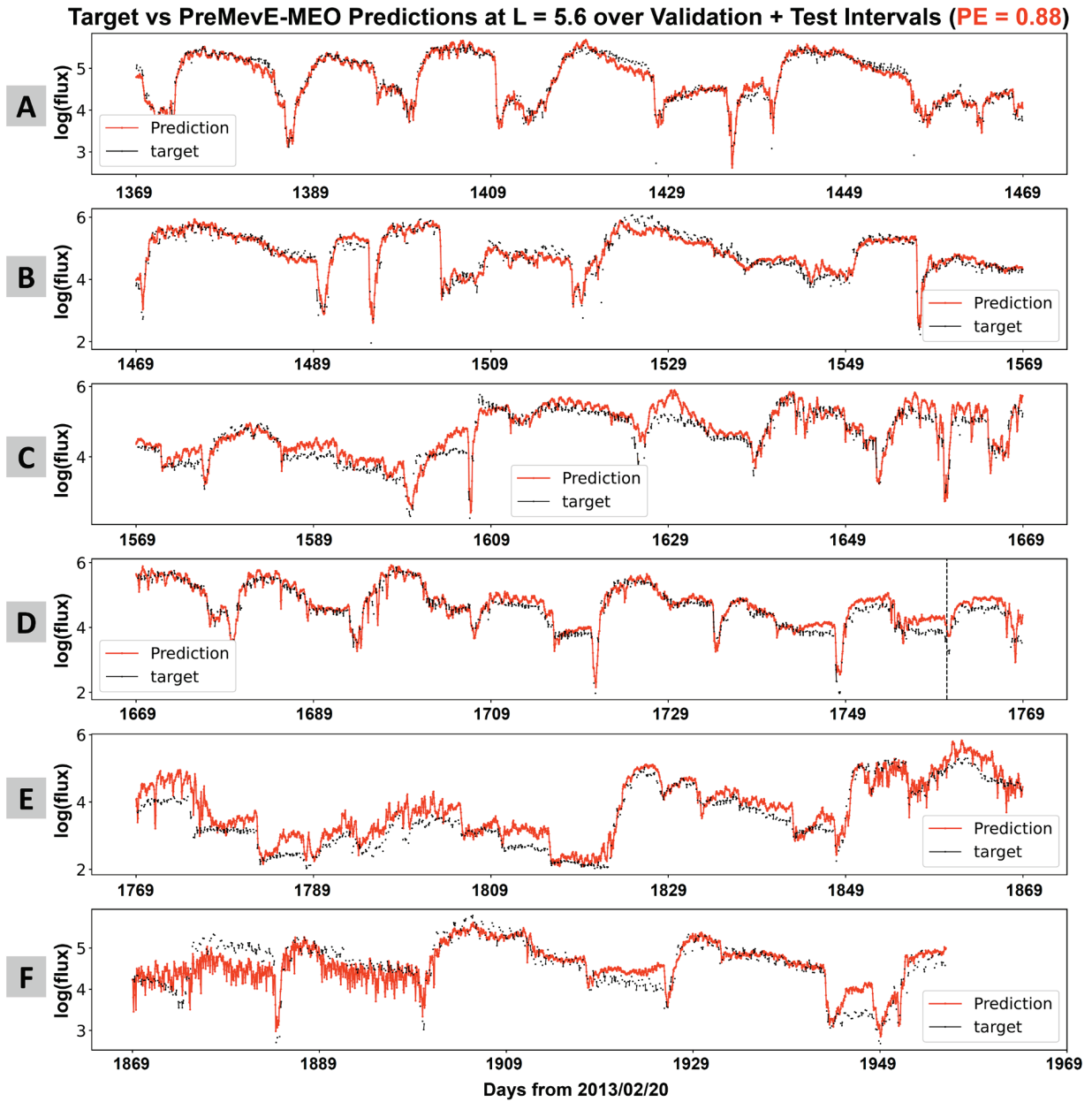

**Figure 6. PreMevE-MEO nowcasts for ≥2 MeV electrons at one L-shell (5.6) over the whole out-of-sample 588 days.** Each panel covers 100 days, in which the target is shown in black and predictions in bright red color.

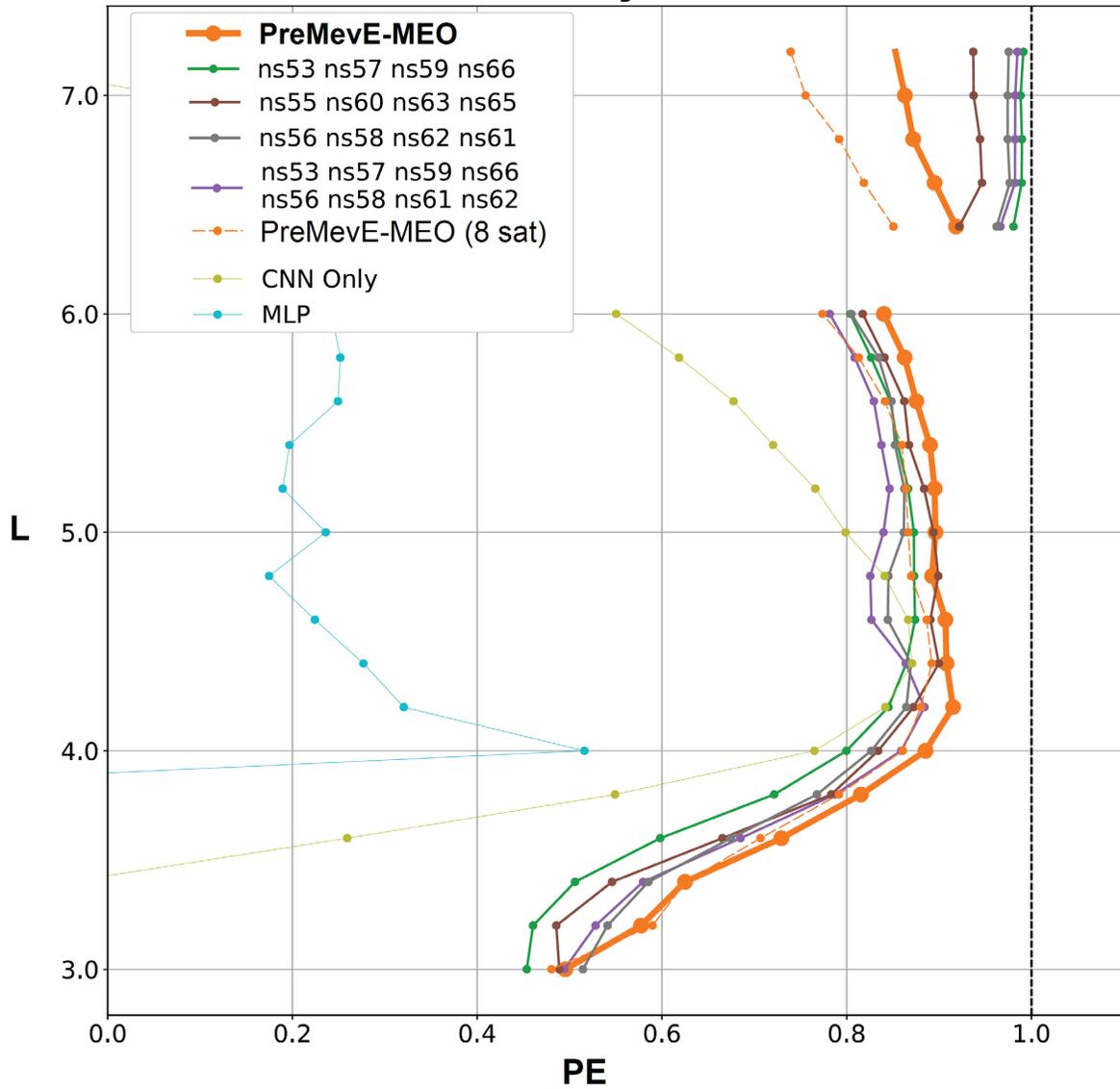

**Figure 7. Model PE values for validation and test data are presented as a function of L-shell for a list of model variations.** The thick orange curve is for the PreMevE-MEO model, compared to models driven by four GPS satellites only (green, brown, and gray), eight satellites retrained (purple), eight satellites non-retrained (thin dashed orange), and inputs from all 12 satellites but other NN algorithms (yellow-green for CNN and cyan for MLP).

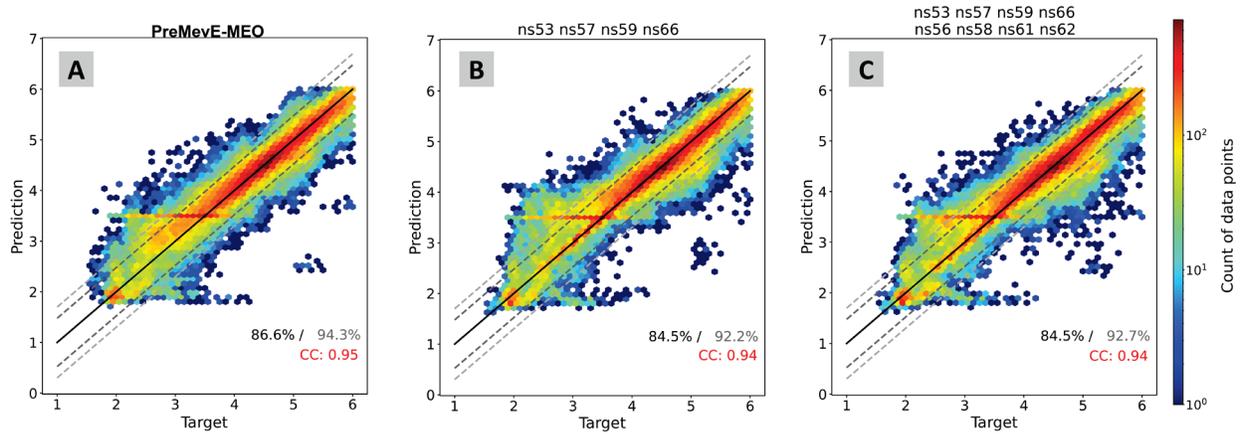

**Figure 8. Model prediction vs. target 2D histograms for nowcasted ≥2 MeV electron fluxes across all L-shells.** A) Histogram of the fluxes predicted by PreMevE-MEO vs. the target ≥2 MeV electron fluxes. Color indicates the count of points in bins of size 0.1 x 0.1. Similarly, panels B) to (D) show predictions vs ≥2 MeV target for one four-satellite input model and one eight-satellite input model. In each panel, diagonal line for perfect matching is shown in solid black curve, and the dashed dark gray (and light gray) lines indicate ratio—between predicted and observed fluxes—factors of 3 (and 5). The dark gray (light gray) number in lower right is the percentage of points falling within the factors of 3 (5), and the red number shows the correlation coefficient value.

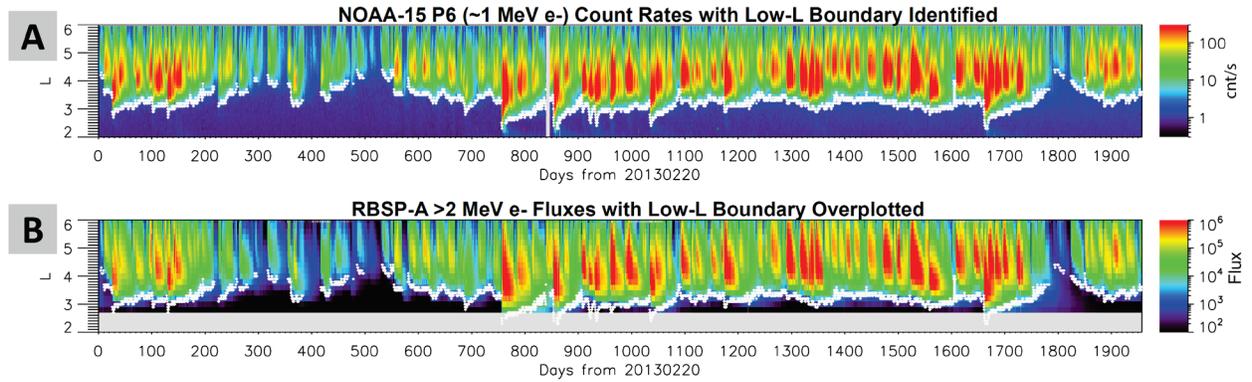

**Figure A1. Low-L boundaries are identified from LEO electron observations. A)** NOAA-15 P6 flux distributions are sorted by L-shell and time over the whole 1957-day interval, with the identified low-L boundary locations overplotted in white. **B)** RBSP ≥2 MeV electron flux distributions are plotted in the same format, overplotted with the same low-L boundary identified from NOAA-15 data.